\documentclass[journal=jacsat,manuscript=article]{achemso}
\usepackage[version=3]{mhchem}   
\usepackage[utf8]{inputenc}
\usepackage[T1]{fontenc}
\usepackage{graphicx}            
\usepackage{dcolumn}             
\usepackage{bm}                  
\usepackage{mathptmx}
\usepackage{amsmath}
\usepackage{gensymb}
\usepackage[normalem]{ulem}
\usepackage{xcolor}
\usepackage{hyperref}            
\usepackage{caption}
\usepackage{subcaption}
\usepackage{wasysym}
\usepackage{physics} 
\usepackage{float}
\usepackage{lipsum} 
\usepackage[mathscr]{eucal}
\usepackage[mathlines]{lineno}   
\author{\textcolor{black}{
    Russell Berger\textsuperscript{1,*}, Alex Mavian\textsuperscript{1}, Edgar Dimitrov\textsuperscript{2}, Na Zhang\textsuperscript{2},\\
    Nazifa Rumman\textsuperscript{3,4}, Pascal Bass\`ene\textsuperscript{1,4,$\dagger$}, Humberto Terrones\textsuperscript{1}, 
    Esther A. Wertz\textsuperscript{1}, Mauricio Terrones\textsuperscript{2}, Edwin Fohtung\textsuperscript{1,5}, 
    and Moussa N'Gom\textsuperscript{1,4,$\ddagger$,}
         }} 
\affiliation{
\textsuperscript{$1$} \footnotesize{Department of Physics, Applied Physics, and Astronomy, Rensselaer Polytechnic Institute, Troy, NY 12180, USA.}\\
\textsuperscript{$2$} \footnotesize{Department of Physics, The Pennsylvania State University, University Park, Pennsylvania 16802, USA.}\\
\textsuperscript{$3$} \footnotesize{Department of Electrical, Computer, and Systems Engineering, RPI, Troy, NY, USA.}\\
\textsuperscript{$4$} \footnotesize{The Shirley Ann Jackson, Ph.D. Center for Biotechnology and Interdisciplinary Studies at RPI, Troy, NY, USA.}\\
\textsuperscript{$5$} \footnotesize{Department of Materials Science and Engineering, Rensselaer Polytechnic Institute, Troy, New York 12180, USA}\\
\textsuperscript{$\dagger$} \footnotesize{Center for Materials, Devices, and Integrated Systems (CMDIS) 6015 Low Center, RPI, Troy, NY, USA.}\\
\textsuperscript{$\ddagger$} \footnotesize{Center for Ultrafast Optical Science, University of Michigan, Ann Arbor, MI 48109-2102}\\
  }
\email{ngomm@rpi.edu ; berger2@rpi.edu}

\title{\Large{Enhancement of Second Harmonic Generation in Monolayer \ce{WS2} by Feedback-Based Wavefront Shaping}}

\abbreviations{2D: Two-Dimensional, WFS: Wavefront shaping,TMD: Transition-Metal Dichalcogenides, SH: Second Harmonic, SLM: Spatial Light Modulator, GB: Grain Boundary , ROI: Region of Interest}

\keywords{Wavefront shaping, Transition-Metal Dichalcogenides, Ultrafast laser, Second harmonic generation, spatial light modulator}

\begin{document}
\setcounter{figure}{-1}

\begin{abstract}
Two-dimensional Transition-Metal Dichalcogenides (TMDs) are of great interest for second harmonic (SH) generation due to their large second-order susceptibility ($\chi^{(2)}$), atomically thin structure, and relaxed phase-matching conditions. TMDs are also promising candidates for miniaturizing nonlinear optical devices due to their versatile applications in photon manipulation, quantum emission and sensing, and nanophotonic circuits. However, their strong SH response is limited by nanometer-scale light-matter interaction and material impurities. Although there is considerable work towards engineering TMDs for enhancing their nonlinear responses, all-optical methods are still in the exploration stages. \\
In this work, we incorporate, to the best of our knowledge, the first experimental demonstrations of feedback-based wavefront shaping (WFS) techniques in atomically thin media to reveal and enhance the weak SH generation of monolayer \ce{WS2}. Phase tuning of the incident wavefront leads to localized regions of high-intensity fundamental light, increasing the intensity of SH generation by up to an order of magnitude in targeted regions. We enhance the local conversion efficiencies from monolayer \ce{WS2} up to 41$\times$ from phase-only modulation. Furthermore, by introducing a shift in the transverse phase structure, we generate observable SH generation at the destructively interfering grain boundaries of polycrystalline monolayers. This method allows for all-optical tuning of TMDs nonlinear responses, opening up possibilities for dynamic signal routing and on-demand enhancement in nanoscale photonic systems.
\begin{figure}[!h]
    \centering
    \includegraphics[width = 0.8\linewidth]{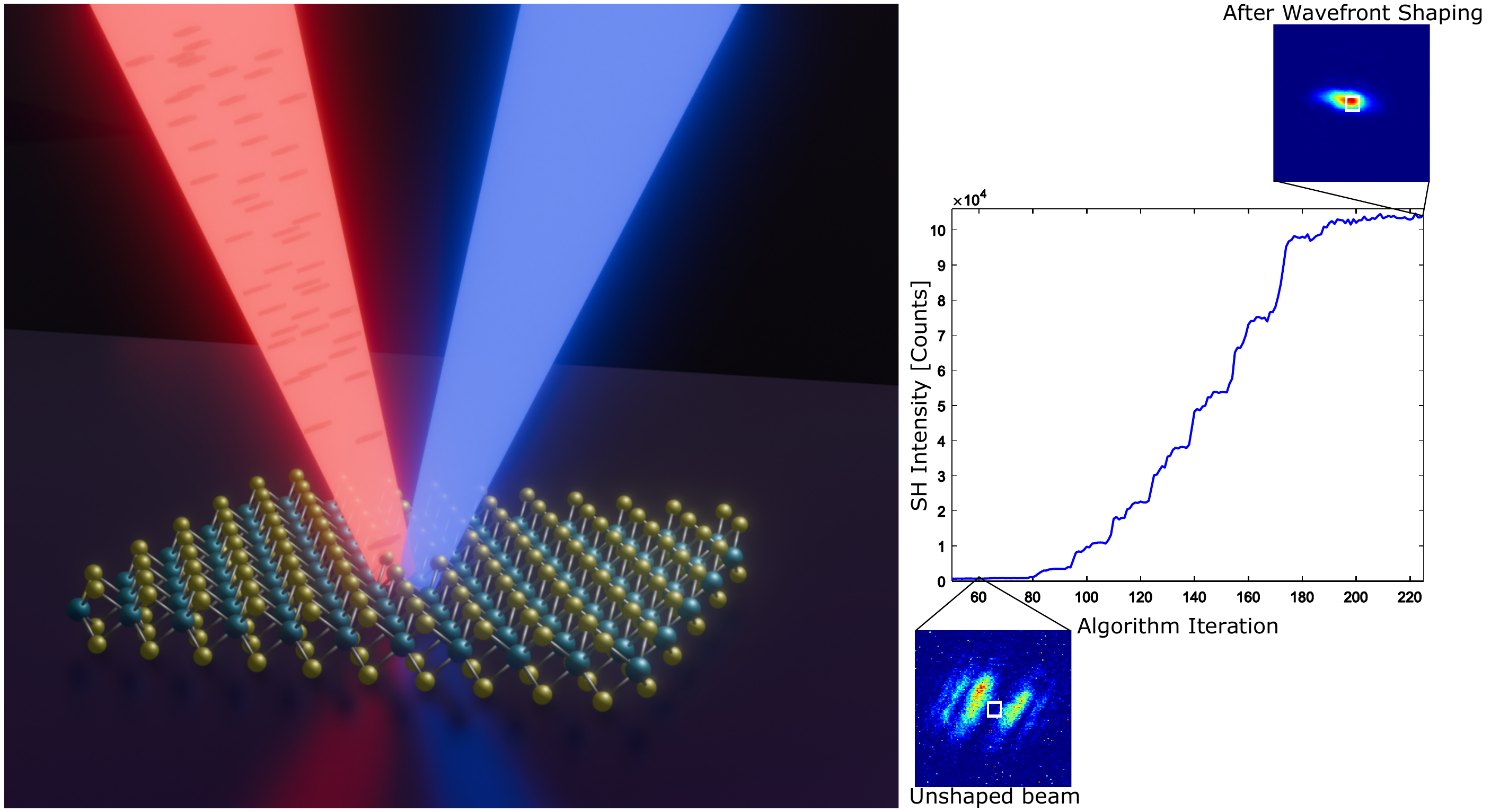}
    \caption{{\bf Illustrative Abstract.} Feedback-based wavefront shaping is used to spatially control and enhance SH intensity from \ce{WS2} monolayers.}
    \label{fig:my_label}
\end{figure}

\end{abstract}

\section{Introduction}

There has been recent interest in two-dimensional (2D) van der Waals materials for their unique properties and their promise for quantum and nanophotonic systems~\cite{doi:10.1021/nn400280c}. Due to their tunable optical and electronic properties, ease-of-integration into photonic circuits, and versatility in photon generation and detection, 2D materials are crucial for the future of quantum and nonlinear devices~\cite{An_Zhao_Zhang_Liu_Yuan_Sun_Zhang_Wang_Li_Li_2022,Ghosh_Dastidar_Thekkooden_Nayak_Praveen_Bhallamudi_2022,Pal_Zhang_Chavan_Agashiwala_Yeh_Cao_Banerjee_2023}. In particular, 2D TMD monolayers have attracted attention for their sizable direct bandgaps, high mechanical stability, great optical and thermal damage thresholds, and low fabrication costs~\cite{CHOI2017116,YouBonguBaoPanoiu+2019+63+97,Kolobov2016}.\\
When scaled down to a single atomic trilayer, TMDs exhibit a high second order optical nonlinearity due to the loss of inversion symmetry, allowing for second order nonlinear optical processes and applications in nanoscale nonlinear photonic circuits~\cite{Autere_Jussila_Marini_Saavedra_Dai_Säynätjoki_Karvonen_Yang_Amirsolaimani_Norwood_et_al._2018,Wen_Gong_Li_2019,Janisch_Mehta_Ma_Elías_Perea-López_Terrones_Liu_2014,Li_Lv_Liu_Yan_Wang_Ren_Wang_Chen_2018}. In particular, it has been found that the $\chi^{(2)}$ for \ce{WS2} and \ce{WSe2} is three orders of magnitude greater than those of conventional bulk nonlinear crystals~\cite{janisch2014extraordinary,Ribeiro-Soares_2015}. In particular, \ce{WS2} has already been implemented as a saturable absorber allowing for stable mode-locking for a pulsed picosecond laser~\cite{Li_Lv_Liu_Yan_Wang_Ren_Wang_Chen_2018}, showing the feasibility of integrating TMDs into nanoscale photonic systems. The atomically thin nature of 2D TMDs allows for relaxed phase-matching conditions, due to the nonlinear interaction occurring at a length much smaller than the coherent interaction length~\cite{Zhang_Zhao_Yu_Yang_Liu_2020,10.5555/1817101}. 

Despite recent advances in tuning and manipulating SH response from TMDs~\cite{Huang_Xiao_Xia_Chen_Zhai_2024}, the short-distance light-matter interaction still hinders SH conversion efficiency, limiting their nonlinear response. Considerable efforts have been made to enhance SH generation in TMDs~\cite{PhysRevLett.114.097403,PhysRevB.87.201401,Seyler_Schaibley_Gong_Rivera_Jones_Wu_Yan_Mandrus_Yao_Xu_2015,doi:10.1021/acs.nanolett.5b02547,Huang_Krasnok_Alú_Yu_Neshev_Miroshnichenko_2022}. Such techniques involve pumping the sample at frequencies corresponding to their exciton resonances~\cite{PhysRevB.87.201401,PhysRevLett.114.097403}, application of external electric fields~\cite{Seyler_Schaibley_Gong_Rivera_Jones_Wu_Yan_Mandrus_Yao_Xu_2015,doi:10.1021/acs.nanolett.5b02547}, integration of the 2D crystals on waveguides, dielectric, and plasmonic nanostructures \cite{Bernhardt_Koshelev_White_Meng_Fröch_Kim_Tran_Choi_Kivshar_Solntsev_2020,Chen_Corboliou_Solntsev_Choi_Vincenti_De_Ceglia_De_Angelis_Lu_Neshev_2017,Spreyer_Ruppert_Georgi_Zentgraf_2021,Wen_Xu_Zhao_Khurgin_Xiong_2018,Shi_Wu_Wu_Zhang_Sui_Du_Yue_Liang_Jiang_Wang_et_al._2022,Li_Wei_Song_Huang_Wang_Liu_Xiong_Hong_Cui_Feng_et_al._2019}, or applying strain to the crystal~\cite{Liu_Yildirim_Blundo_de_Ceglia_Khan_Yin_Nguyen_Pettinari_Felici_Polimeni_et_al._2023}. While most works present significant engineering efforts in tuning the nonlinear optical response of TMDs, all-optical methods remain relatively scarce.
Spatial light modulators (SLMs) allow for novel physics due to the phase tunability of the many transverse modes of an incoming beam~\cite{Maurer_Jesacher_Bernet_Ritsch-Marte_2011}. The ability to manipulate light~\cite{N’Gom_Lien_Estakhri_Norris_Michielssen_Nadakuditi_2017} has applications in biological imaging~\cite{Yu_Park_Lee_Yoon_Kim_Lee_Park_2015,Wang_Rumman_Bassène_N’Gom_2023,Rumman_Wang_Jennings_Bassène_Buldt_N’Gom_2022} and signal transmission~\cite{N’Gom_Norris_Michielssen_Nadakuditi_2018,Wang_Ali_Reza_Buldt_Bassène_N’Gom_2023,AliReza:23,Tran_Wang_Nazirkar_Bassène_Fohtung_N’Gom_2023}. Also, it has been demonstrated that through the use of an SLM, one can improve the SH generation efficiency by an order of magnitude in a type-II KTP nonlinear crystal by introducing a transverse phase gradient~\cite{Thompson_Hokr_Throckmorton_Wang_Scully_Yakovlev_2017}. SLMs have also been used to improve SH generation efficiency through nonlinear crystal powders and nonlinear scattering media~\cite{Qiao_Peng_Zheng_Ye_Chen_2017,Wu_Fan_Chen_Pu_2022}. All these recent works focus on SH generation in media where phase-matching conditions are non-negligible. However, to the best of our knowledge, WFS experiments on SH signal optimization in atomically thin media have yet to be explored.

In this work, we demonstrate the enhancement of SH response in  chemical vapor deposition (CVD) grown monolayer \ce{WS2} under ambient conditions through phase-only modulation of the pump beam. The wavefront of the incident beam is shaped with an SLM using feedback-based wavefront shaping techniques~\cite{Vellekoop:15,Wang_Rumman_Bassène_N’Gom_2023} to enhance the nonlinear response in targeted regions for each \ce{WS2} sample.  We examine the SH generation in both uniform and polycrystalline monolayer \ce{WS2}, with similar experiments performed on multi-layer \ce{WS2} [SI, Sec.~4, Figure~S5]. The shaped phase of the incident beam creates localized areas of enhanced SH response. The degrees of freedom in the phase mask of SLMs allow for high tunability and control of the SH intensity at most desired spatial locations from TMDs. Our results show that WFS can be used to overcome the aforementioned weak interaction by endowing the pump with a phase necessary for optimal SH generation.  Furthermore, polycrystalline TMD monolayers form interfaces known as grain boundaries (GBs). 
At the GBs of monolayer polycrystals, neighboring grain sites are misaligned 60$\degree$ with respect to each other. This creates a spatially dependent geometric phase difference which results in destructive interference of SH light generated at the GB~\cite{Yin_Ye_Chenet_Ye_O’Brien_Hone_Zhang_2014,Dasgupta_Yang_Gao_2020}. These diminish not only the intensity of the emitted SH light but also impose a restriction on the spatial regions where SH light can be produced. By compensating for this phase offset via the SLM, we show that SH light can in fact be generated at these GBs. The ability to use feedback-based optimization algorithms allows for on-demand tuning and dynamic signal routing of the nonlinear optical response, opening up new avenues for applications in nanophotonics, materials characterization, and nonlinear microscopy.

\section{Experiment}

\begin{figure}[!t]
    \centering
    \includegraphics[width = \textwidth]{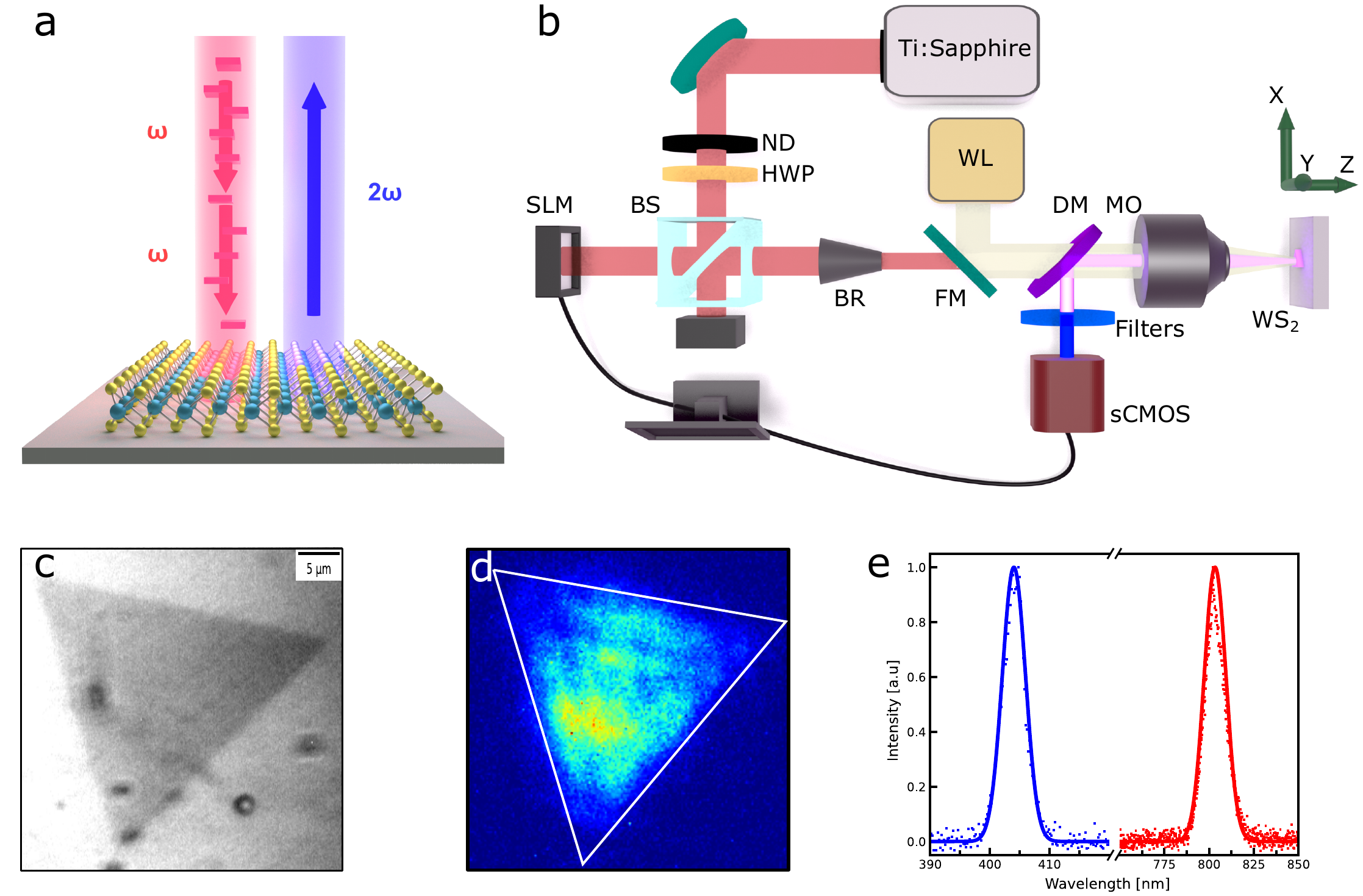}
    \caption{{\bf Concept and schematic of the experimental setup.}~a) Schematic to illustrate SH response enhancement via shaping the wavefront of the fundamental beam.~b) Diagram of the optical setup; ND: neutral density filter, HWP: half waveplate, SLM: spatial light modulator, BS: beam splitter, BR: beam reducer, FM: flippable mirror, WL: white light source DM: dichroic mirror, MO: microscope objective (50$\times$, NA~$=$~0.55).~c) Image of a \ce{WS2} monolayer illuminated by the white light source.~d) Corresponding SH image of the sample in c).~e) Normalized spectra of the SH signal and incident fundamental beam.}   
    \label{fig:experiment_polarimetry}
\end{figure}

\ce{WS2} samples are synthesized using CVD on a 285~nm \ce{SiO2}/\ce{Si} substrate. Further details on the sample preparation and growth are presented in [SI, Sec.~1]. To determine the number of layers, photoluminescence data~\cite{Gutiérrez_Perea-López_Elías_Berkdemir_Wang_Lv_López-Urías_Crespi_Terrones_Terrones_2013}, Raman spectroscopy~\cite{Berkdemir_Gutiérrez_Botello-Méndez_Perea-López_Elías_Chia_Wang_Crespi_López-Urías_Charlier_et_al._2013}, and atomic force microscopy data were taken and examined alongside the optical microscope images~\cite{Benameur_Radisavljevic_Héron_Sahoo_Berger_Kis_2011}. The optical characterization methods are presented in [Si, Sec.~2, Figures~S1~and~S2]. The Samples were illuminated with a Ti:Sapphire femtosecond laser (Spectra Physics Tsunami, 804~nm, 80~MHz, 100~fs), with the experimental setup depicted in Figure~\ref{fig:experiment_polarimetry}.b. The incident Gaussian beam is reflected by the SLM, on which a holographic phase mask is displayed before being focused onto the sample surface by a 50$\times$ microscope objective. Most of the monolayer samples have dimensions of $\sim$20~$\mu$m in length, and $\sim$0.7~nm in thickness. Before measurement, the $z-$position of the sample is adjusted to fully 
 illuminate the entire sample, as shown in Figure~\ref{fig:experiment_polarimetry}.d. 
 
To identify individual flakes, an in-house microscope was constructed with a flippable mirror allowing to switch between white light and laser illumination. Figure~\ref{fig:experiment_polarimetry}.c shows a \ce{WS2} monolayer sample observed using our microscope. 
The generated SH is collected in reflection by the MO as shown in Figure~\ref{fig:experiment_polarimetry}.b, separated from the fundamental by a dichroic mirror (800/400~nm), filtered using a bandpass filters (400~$\pm$~20~nm), before being observed on an sCMOS detector (Thorlabs CS2100M-USB). Figure~\ref{fig:experiment_polarimetry}.d depicts a typical SH image observed on our detector, and Figure~\ref{fig:experiment_polarimetry}.e depicts the measured spectrum, confirming that our measured signal is half the wavelength of our fundamental beam.
The total SH intensity counts within a select a region of interest (ROI) and use the total intensity counts as the feedback parameter for the WFS algorithm. After a measure of the background, the continuous sequential phase-shaping algorithm is used to optimize the SH response. Once the WFS process is completed, the resulting SH signal before and after optimization is analyzed. 

\section{Results and discussion}

\begin{figure}[!t]
  \centering
  \includegraphics[width = \textwidth]{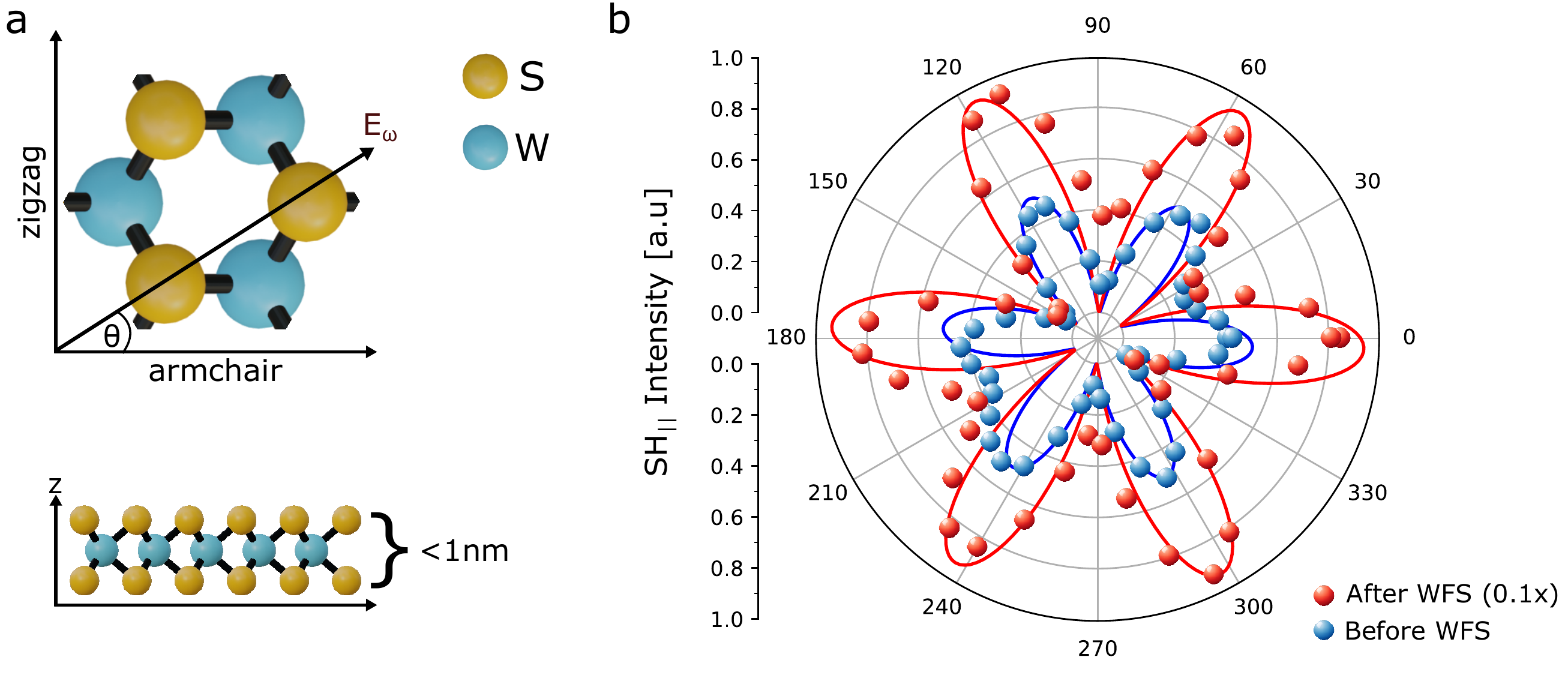}
  \caption{{\bf \ce{WS2} parallel SH polar distribution.}~a) Top view of a lattice structure of \ce{WS2} is shown at the top with the zigzag and armchair directions defined. A side profile of a single atomic three-layer (bottom) is also shown.~b) Polar distribution of the parallel SH intensity before and after WFS as a function of $\theta$, the polarization angle of the fundamental beam. The SH signal after WFS is 10$\times$ scaled down compared with the signal before WFS for better visualization of the polar distribution.}
  \label{fig:Polar}
\end{figure}

\begin{figure}[!t]
  \centering
  \includegraphics[width = 0.9\textwidth]{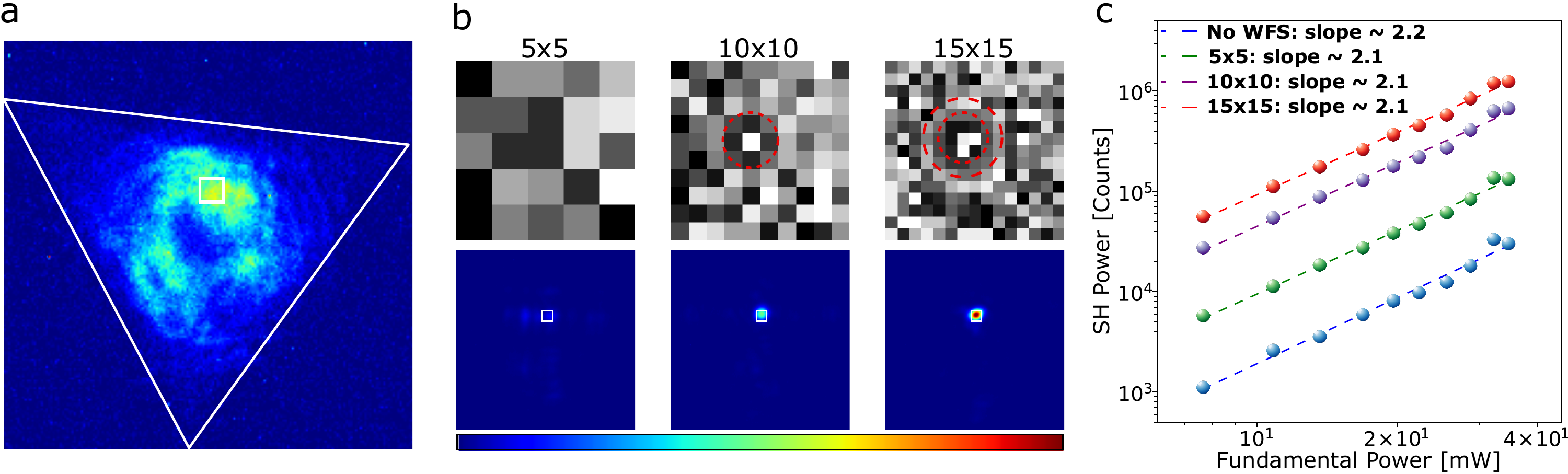}
  \caption{{\bf Effect of phase mask size on SH enhancement.}~(a) SH intensity from \ce{WS2} monolayer. The white triangle follows the shape of the monolayer while the white square inset indicates the target region selected for enhancement.~(b) Optimization phase masks (top) and corresponding SH signal (bottom) with color scale normalized to the maximum SH intensity for the 15$\times$15 phase mask. Dashed red circles indicate regions of phase comparable to that of a digital lens~(c) Slope of the quadratic dependence SH power as a function of fundamental power for each targeted region. The measured power is represented by the points and are fitted by a dash line.}
  \label{fig:CE}
\end{figure}

Single layer \ce{WS2} belongs to the noncentrosymmetric $d_{3h}$ crystal group, with non-zero second-order susceptibility tensor elements $\chi^{(2)}_{yyy}=-\chi^{(2)}_{yxx}=-\chi^{(2)}_{xyx}=-\chi^{(2)}_{xxy}$. Figure~\ref{fig:Polar}.a shows both the hexagonal crystal structure of \ce{WS2} from a top-down view, and the thickness-scale of a single TMD layer viewed from the side. $\theta$ represents the angle between the direction of the incident beam polarization and the armchair direction~\cite{Kumar_Najmaei_Cui_Ceballos_Ajayan_Lou_Zhao_2013}. The threefold rotational symmetry of the $d_{3h}$ group allows for a distinct sixfold polarized SH response when examining either the parallel or the perpendicular component of the SH light compared to the incident polarization. Thus, SH generation can be used to determine the crystal orientation for individual and layered samples~\cite{Hsu_Zhao_Li_Chen_Chiu_Chang_Chou_Chang_2014,Psilodimitrakopoulos_Mouchliadis_Paradisanos_Lemonis_Kioseoglou_Stratakis_2018,Psilodimitrakopoulos_Mouchliadis_Paradisanos_Kourmoulakis_Lemonis_Kioseoglou_Stratakis_2019}.\\
To further confirm the nonlinear character of the generated response imaged and spectrally measured in Figures~\ref{fig:experiment_polarimetry}.(d, e), we examine its polar distribution. The parallel component of the SH intensity (I$_{||}$) follows the expression~\cite{10.5555/1817101} $ I_{||} \propto \cos^2[3(\theta-\phi)] $ where $\phi$ represents the initial polarization angle of the fundamental beam. \\
For all of our studies, the intensity of the SH beam is defined as the total counts within the target region. As shown in Figure~\ref{fig:Polar}.b, the initial SH response clearly exhibits the expected sixfold symmetry. After WFS, we observe the same sixfold distribution, enhanced over an order of magnitude. Note that the SH intensity distributions around 30$\degree$ and 210$\degree$ before WFS do not exhibit much intensity contrast relative to their adjacent intensity maxima ([359$\degree$, 59$\degree$] and [179$\degree$, 239$\degree$], respectively). Slight deviations from theoretical predictions can be attributed to potential defects within the \ce{WS2} crystal. We notice that after WFS, the same range of polarization angles exhibits much greater contrast with respect to their maxima.\\
In previous works\cite{Thompson_Hokr_Throckmorton_Wang_Scully_Yakovlev_2017}, it has been shown that the phase mask resolution applied to the SLM influences the amplitude of the generated SH signal. This is due to the transverse phase gradient affecting the generated SH amplitude. Here, we examine whether this applies to nanometer-scale interaction lengths, where the phase-matching condition is relaxed. 
A SH image of the monolayer \ce{WS2} used for this experiment is shown in Figure~\ref{fig:CE}.a. In Figure~\ref{fig:CE}.b is shown the resulting optimized SH signals, and their respective phase masks displayed above them. We can see that for the 5x5 phase mask, the SH response is more apparent within the ROI, yet some signal is still observed outside the target region. 
\\
In our experiments, the optimization algorithm only seeks to maximize the signal within the selected ROI and does not consider the resulting signal outside the target region. Other algorithms have been developed to allow further degrees of freedom in signal optimization~\cite{Wang_Rumman_Bassène_N’Gom_2023}, which could potentially minimize the signal outside the desired target area, but they are not used for these experiments. As the phase mask resolution increases, the SH signal within the target area increases. 
For the 10$\times$10 and 15$\times$15 phase mask, the optimized SH response is focused tighter than that of the 5$\times$5 phase mask. Between the 10$\times$10 and 15$\times$15 mask, we observe a region of phase resembling to that of a lens~\cite{Rosales-Guzman_Forbes_2017}. These regions are marked with red dashed circles to indicate superpixels of comparable phase. \\
Further details on the comparison between WFS technique and a digital lens displayed onto the SLM, can be found in [SI, Sec.~3, Figure~S3]. In the case of the digital lens, prior information about the focus distance, beam wavelength, and spatial offset are needed to fine-tune the response. 
With feedback based WFS, not only are we able to obtain similar results, but we are also able to optimize several areas at once [SI, Sec.~5, Figure~S6]

Next, the dependence of SH signal power with respect to the fundamental excitation power is characterized. The SH power is expected to scale quadratically with the fundamental power via the relation $ P_{2\omega} \propto \eta P_{\omega}^2 $, where $\eta$ represents the relative conversion efficiency of the TMD. To vary the fundamental power, a variable ND filter is placed in the fundamental beam path (Figure~\ref{fig:experiment_polarimetry}). The power of the fundamental beam is measured after the SLM. It has shown no variation before and after WFS, indicating that only the phase of our pump beam is modified. Figure~\ref{fig:CE}.c shows the effects of WFS on the local SH conversion efficiency for the \ce{WS2} sample. The dashed fit lines in the log-log plot demonstrate the expected quadratic behavior of the nonlinear process for each phase mask. We define the conversion efficiency enhancement factor ($\zeta$) as the ratio $\zeta = \eta_{WFS}/\eta$, where $\eta_{WFS}$ is the SH generation efficiency after optimization. In Figure~\ref{fig:CE} for 5$\times$5, 10$\times$10, and 15$\times$15 phase masks, we report $\zeta$ to be 4.41, 21.19, and 41.16, respectively, indicating an 
\begin{figure}[!t]
 \centering
 \includegraphics[width = 0.9\textwidth]{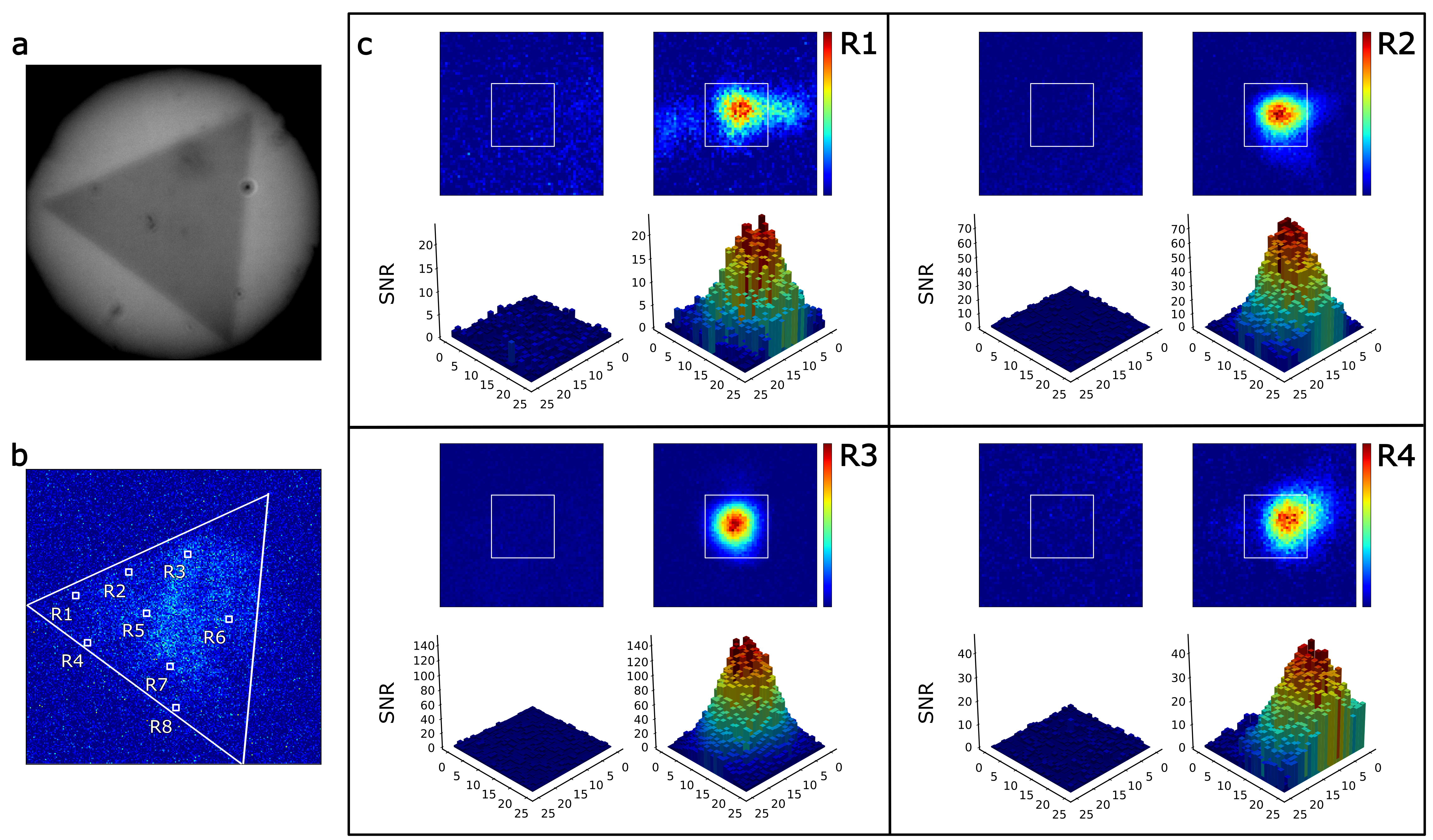}
 \caption{{\bf Intensity enhancement and spatial target selection.}~(a) Optical microscopy image of \ce{WS2} monolayer under white source illumination.~(b) Corresponding SH intensity distribution before the optimization process. The white box insets indicate each target region R$_i$ ($i\in[1, 8]$).~(c) The SH intensity distribution (top) and (SNR), Signal-to-noise ratio (bottom) are shown before (left) and after (right) WFS. Only R$_1-$R$_4$ target regions are shown for more clarity.}
   \label{fig:ML}
\end{figure}\\
order of magnitude enhancement in the generated SH signal. During the WFS process, the phase modulation changes the focus of the fundamental light on the sample. Therefore, the incident electric field (intensity) increases locally, enhancing the efficiency factor without changing the overall power.   

To demonstrate the spatial flexibility of our method, we examine 8 different target regions across a single \ce{WS2} monolayer triangle. Figure~\ref{fig:ML}a shows the optical image of the sample we are examining, and Figure~\ref{fig:ML}b shows the corresponding SH intensity distribution before optimization. In this run, the sample is kept stationary, and only the choice of the target region is changed between each run. A 5$\times$5 pixel target area is chosen, and a 15$\times$15 phase mask is used in each run. The SH intensity distribution before and after WFS, along with the corresponding signal-to-noise ratio (SNR) map is shown for a chosen 4 target regions in Figure~\ref{fig:ML}c (see SI, Sec.~4, Figure~S4 for the rest of target areas). The enhancement factor $\gamma$ is a ratio $\langle I_{WFS}\rangle/\langle I_0\rangle$ between I$_{WFS}$ and I$_0$, the  average intensity in the ROI after and before optimization. For regions R$_1-$R$_4$, we report $\gamma$ = 25.1, 86.2, 101.0, and 21.9, respectively, demonstrating one to two orders of magnitude enhancement. We note greatly increased signal intensity within each of the chosen target areas, allowing for localized enhancement of the SH response at several locations. Once an optimal phase mask has been determined, we are able to freely switch between different masks for each target area, allowing for dynamic control of the SH generation.\\
The SNR for each target region is determined by comparing the signal intensity within the target region divided by the background standard deviation of the detector. Before optimization, some regions have SH signals that are indistinguishable from the background noise with an SNR close to 1. After optimization, the generated SH intensity can reach SNR values twice the order of magnitude before WFS (Figure~\ref{fig:ML}c, region 3). This method demonstrates great intensity enhancement of generated SH signal from \ce{WS2} single layer thus improving the feasibility and consideration of \ce{WS2} as a low dimensional nonlinear photonic device.

In CVD-grown monolayer \ce{WS2}, polycrystals can be formed~\cite{Artyukhov_Hu_Zhang_Yakobson_2016,Lan_Zheng_Ding_Hong_Wang_Li_Li_Yang_Hu_Pan_et_al._2024}, which possess grain boundaries (GBs) that result in weak or no SH signal. Recently, it has been suggested that the reason for the such low SH generation at the grain boundaries is due to the geometric phase superposition in the generated SH signal. Neighboring crystals are oriented 60$\degree$ with respect to each other~\cite{Dasgupta_Yang_Gao_2020} and that misorientation often observed in star-shaped and bowtie-shaped monolayers~\cite{Artyukhov_Hu_Zhang_Yakobson_2016} induces a $\pi-$phase shift between the neighboring SH signals~\cite{Dasgupta_Yang_Gao_2020}, destructively interfering at the GB.
\begin{figure}[!t]
    \centering
    \includegraphics[width = \textwidth]{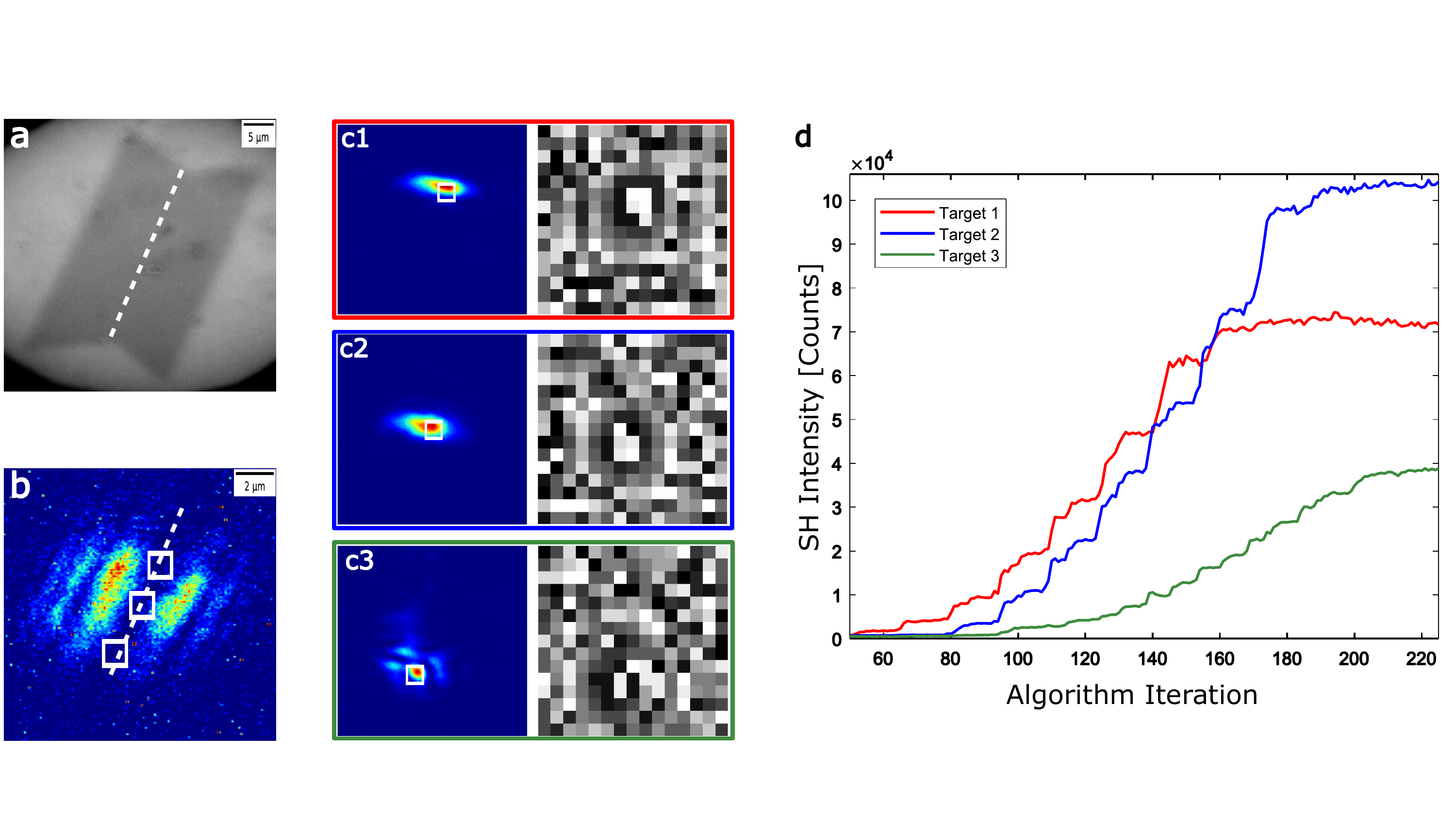}
    \caption{{\bf WFS for spatial control of SH generation.}~(a) Optical image of \ce{WS2} a bowtie  sample. The white dashed line indicates the GB where SH is expected to destructively interfere~(b) Far field SH intensity distribution before WFS. White box insets indicate target regions selected for feedback.~(c1--c3) Optimized SH intensity distribution and the final phase mask for each WFS run.~(d) SH Intensity enhancement within each target area as a function of the number of iterations. 
    }
   \label{fig:GB}
\end{figure}\\
Figure~\ref{fig:GB}a shows an optical image of such a bowtie sample, along with its far field SH intensity distribution in Figure~\ref{fig:GB}b. In the SH image, we can clearly remark the presence of a grain boundary connecting the reentrant points of the TMD sample, creating a symmetric-like shape in the SH intensity image. This region exhibits destructive interference in the SH generation, consistent with the suggested idea that the neighboring crystals are misoriented with respect to each other. For the optimization process, we use the far field SH intensity map, and choose target regions in between the bright SH signals as indicated by the white boxes in Figure~\ref{fig:GB}b. Since the diminished SH signal is due to a phase difference between the neighboring crystals, our technique provides an avenue to mitigate this destructive interference. \\
By adjusting the phase of the beam, an additional phase offset occurs prior to the fundamental interaction with the TMD crystal, compensating for the destructive interference observed at the grain boundary. Figures~\ref{fig:GB}c1$-$c3 show the resulting SH signal after the optimization process and the final phase mask profiles. Figure~\ref{fig:GB}d shows the SH intensity evolution as a function of the number of iterations. Here, an iteration refers to a generated superpixel. We observe that the final SH intensity for each run differs significantly and is correlated the position of the target region. To generate SH light within the target regions, the local phase corresponding to the fundamental excitation within that region must be adjusted, and by modifying the phase of the fundamental beam, we are able to overcome the intrinsic properties of polycrystalline \ce{WS2} and generate SH light in regions previously thought to be hindered by grain boundaries.  

\section*{Conclusion}

In summary, we have experimentally demonstrated an all-optical,  targeted enhancement in the nonlinear response of monolayer and polycrystalline \ce{WS2} fabricated using CVD. The optimized wavefront allows for the fundamental light to create localized focus points, which allows more of the energy to be pumped into the SH field. We have shown the potential of our method for enhanced local conversion efficiency and dynamic signal routing of second harmonic response. Our results show that the phase modulation of the incident light can lead to a large improvement of the nonlinear response of nanoscale materials. \\
This work can be combined with wavelength and polarization tuning, along with introducing orbital angular momentum~\cite{Dasgupta_Gao_Yang_2019} to greatly increase the degrees of freedom for information transmission using SH light. Unlike other SH enhancement techniques~\cite{doi:10.1021/acs.nanolett.5b02547,Huang_Krasnok_Alú_Yu_Neshev_Miroshnichenko_2022,PhysRevB.87.201401,PhysRevLett.114.097403,Seyler_Schaibley_Gong_Rivera_Jones_Wu_Yan_Mandrus_Yao_Xu_2015,Bernhardt_Koshelev_White_Meng_Fröch_Kim_Tran_Choi_Kivshar_Solntsev_2020,Chen_Corboliou_Solntsev_Choi_Vincenti_De_Ceglia_De_Angelis_Lu_Neshev_2017,Liu_Yildirim_Blundo_de_Ceglia_Khan_Yin_Nguyen_Pettinari_Felici_Polimeni_et_al._2023}, our method is realized under ambient room conditions with samples not mechanically or chemically modified. These results indicate the possibility of using wavefront shaping to create on-demand enhancement of selected regions in low-dimensional nonlinear media. 


\begin{suppinfo}
Supporting information is available in the accompanying pdf.
\begin{itemize}
    \item Further details on sample preparation; Raman, photoluminescence and AFM characterization of \ce{WS2} samples(Figures~S1-S2); comparison between feedback-based wavefront shaping and digitals lens phase masks(Figure~S3); additional SNR maps from monolayer \ce{WS2} (Figure~S4); Additional results on multilayer and star-shaped polycrystalline \ce{WS2} (Figures~S5-S6).
\end{itemize}
\end{suppinfo}
\section*{Author Contributions}
R.B. conceived and performed the design. R.B. and A.M. performed optical measurements and analyzed data. E.D. and N.Z. fabricated samples under M.T.'s supervision.  N.R. wrote the wavefront shaping algorithm used in this work. M.N'G. and P.B. organized and led the project. Raman and AFM measurements were performed by R.B. under H.T.'s supervision. M.N'G., E.W., and E.F. supervised writing and revision of the manuscript. M.N'G. provided funding for the experiment.\\
R.B. wrote the paper, with input from all authors. All authors discussed the results.

\section*{Notes}
The authors declare no conflict of interest.

\section*{Funding}
 P.B., E.F., and M.N’G. acknowledge the support by a grant
 from the Gordon and Betty Moore Foundation to the PAIR
UP Imaging Science Program. 
M.N’G. acknowledges the support of the U.S. Department of Energy
 (DOE) under Grant No DE-SC0024676. 
 M.N’G. acknowledges the support from the National Geospatial Intelligent Agency Grant No. HM04762010012.
E.F. acknowledges support from the US Department of Energy (DOE), Basic Energy Sciences, Materials Sciences and Engineering Division, under grant No. DE-SC0023148 and Air Force Office of Scientific Research (AFOSR), under award No. FA9550-23-1-0325. The authors acknowledge the use of the VESTA 3 software for generation of \ce{TMD} models~\cite{Momma:db5098}.

\bibliography{References}

\end{document}